\documentclass[graybox]{svmult}

\usepackage[utf8]{inputenc}
\usepackage[ukrainian,english]{babel}

\usepackage{mathptmx}       
\usepackage{helvet}         
\usepackage{courier}        
\usepackage{type1cm}        
\usepackage{makeidx}         
\usepackage{graphicx}        
\usepackage{multicol}        
\usepackage[bottom]{footmisc}

\usepackage[normalem]{ulem}

\RequirePackage{color}
\definecolor{MyDarkGreen}{rgb}{0.02,0.60,0.06}

\makeindex             

\begin{document}

\title*{Complex Networks of Words in Fables}
\author{Yurij Holovatch and Vasyl Palchykov}
\institute{Yurij Holovatch \at Institute for Condensed Matter Physics, National Acad. Sci. of Ukraine, 79011 Lviv, Ukraine,
\email{hol@icmp.lviv.ua}
\and Vasyl Palchykov \at Institute for Condensed Matter Physics, National Acad. Sci. of Ukraine, 79011 Lviv, Ukraine,
\at Lorentz Institute for Theoretical Physics, Leiden University, 2300 RA Leiden, The Netherlands, \email{palchykov@icmp.lviv.ua}}
\maketitle

\abstract{In this chapter we give an overview of the application of
complex network theory to quantify some properties of language. Our
study is based on two fables in Ukrainian, {\em Mykyta the Fox} and
 {\em Abu-Kasym's slippers}. It consists of two parts: the analysis
of frequency-rank distributions { {of words}} and the application of
complex-network theory. The first part shows that the text sizes are
sufficiently large to observe statistical properties. This supports
their selection for the analysis of typical properties of the
language { {networks}} in the second part of the chapter. In
describing language as a complex network, while  words are usually
associated with nodes, there is more variability in the choice of
links and different representations result in different networks.
Here, we examine a number of such representations of the language
network and perform a comparative analysis of their characteristics.
Our results suggest that, irrespective of link representation, the
Ukrainian language network used in the selected fables  is a
strongly correlated, scale-free, small world. We discuss how such
empirical approaches  may help form a useful basis for a theoretical
description of language evolution and how they may be used in
analyses of other textual narratives.}

\section{Introduction}
\label{sec:1} Applications of  methods of quantitative analysis that
are widely used in natural sciences gave rise to the discovery of
one of the best known empirical relationships  of quantitative
linguistics. In its simplest form, this states that the probability
to randomly select the $r$-th  most frequent word in a text is $r$
times smaller than the probability to randomly select the most
frequent one (Zipf 1935, Zipf 1949). Representing the probability to
randomly select the $r$-th  most frequent word as $f(r)$, this
relation may be expressed by the  equation
\begin{equation}\label{eq1.2}
f(r) = A/r^{\alpha},
\end{equation}
where $\alpha=1$. The empirical { {observation}} that $\alpha $ has
the same value (namely 1) for so many natural language utterances
is what is remarkable about this equation.  (The quantity $A$ is
less remarkable, being just a normalization coefficient which
ensures the total probability properly sums to one). It means the
distributions are ``fat-tailed'' -- characterised by rare events
happening more frequently than for normal distributions. This
discovery is often attributed to its populariser, Harvard
linguistics professor G.~K.~Zipf (Zipf 1935), however,
 similar observations have been reported previously by J.~B.~Estoup
(1916) and E.~U.~Condon (Condon 1928). Some deviations from the
original form of  Zipf's law (\ref{eq1.2}) were later observed
(Kanter et al. 1995, Montemuro 2001), but the fat tail of  the
distribution, which is a typical signature of the long-range
correlations between words within corpora, remained unchanged.
Moreover, analysis of vast corpora lead to a conclusion about two
scaling regimes characterizing the word frequency distributions,
with only the more common words (the so called kernel or core
lexicon) obeying the classic Zipf law (Ferrer i Cancho and Sol\'e
2001a and Petersen et al. 2012). While {\em local} organization of
words within single sentences appears to be quite natural, due to
the rules of grammar, long-range correlations between words are far
from trivial (Kanter et al. 1995). Nonetheless, why such local
sequences become organized {\em globally} may be explained by
various mechanisms (Simon 1955, Li 1992).

Zipf's law and the reasons behind it provide only superficial
understanding of the organization of language since the connections
between the words are neglected. These connections reflect the
organization of words into sentences and play  key roles for
transmitting  information (Ferrer i Cancho and Sol\'e 2001b). Thus,
to investigate deeper structural characteristics of a language the
relationships between the language units (such as words) should be
taken into account. A set of those relationships combined with the
corresponding language units may be naturally represented as a
network or a graph. Such representations allow one to apply a number
of tools to investigate the properties of the underlying system on
various scales (Newman 2010).

In seeking to represent a language by a network,  one may decide to
relate words syntactically (Ferrer i Cancho et al. 2004, Ferrer i
Cancho et al. 2005, Sol\'e 2005, Masucci and Rodgers 2006, {
{Corominas Murtra et al. 2007, Sol\'e et al. 2010, Barcel\'o-Coblijn
et al. 2012}}) or semantically (Motter et al. 2002, Sigman and
Cecchi 2002, de Jesus Holanda et al. 2004, Borge-Holthoefer and
Arenas 2010, { {Sol\'e and Seoane 2014}}), for example.
Alternatively one may chose to link words together based on their
co-occurrence (i.e., if they appear adjacent to each other) or on
their having appeared in the same sentence (Ferrer i Cancho and
Sol\'e 2001b, Caldeira et al. 2006, Holovatch and Palchykov 2007,
Zhou et al. 2008, { {Sol\'e and Seoane 2014}}). Therefore, there is
no unique network representation of a language. Nonetheless many
features of language networks are shared, not only among various
representations, but for diverse languages as well (Ferrer i Cancho
et al. 2004). The key common features of these networks include the
{\emph{small world}} structure (Watts 1999) and the
{\emph{scale-free}} topology (Albert et al. 1999), both
characterizing the global picture of linking architecture. The
former (the small world effect) demonstrates that the links connect
the words in a specific way that makes the corresponding structures
extremely compact. The latter  (the scale-free topology)
demonstrates that the number of links that are connected to an
arbitrarily selected node in a network have extremely high levels of
fluctuations. The number of links of a given node is called its
{\emph{degree}} and is represented by the variable $k$. The
scale-free property is then represented mathematically by a degree
distribution function $P(k)$ which has a power-law decay as follows:
\begin{equation}\label{eq1.1}
P(k)\sim k^{-\gamma},
\end{equation}
in which the exponent $\gamma>1$. (Here the symbol $\sim$ means
``behaves asymptotically like'', having suppressed a normalisation
term.) These structural peculiarities reveal high level of system
heterogeneity and have to be properly considered in investigations
of the  mesoscopic structure of human language (Newman 2012).

In this article we report on an analysis of the Ukrainian language
through two representative texts, namely  {\em Mykyta the Fox} and
{\em Abu-Kasym's slippers} (Holovatch and Palchykov 2007). The
former was written by the prominent Ukrainian writer Ivan Franko
and the latter adapted by him into Ukrainian. For each fable, and
for a combination of the two, we construct several network
representations, where the links between nodes (words) are
introduced in various ways based on the interaction window. The
properties of the different network representations are compared to
each other and  features of the Ukrainian language are compared to
ones of other well studied languages.

The fable {\em Mykyta the Fox} is based on a story about a clever
Fox and his adventures in the kingdom of a Lion. {\em Abu-Kasym's
slippers } is a story about a miserly merchant in Baghdad. One may
find variants of this  story in many cultures. Therefore, similar to
other chapters of this book, we use written narratives for our
analysis. However, here we are interested in universal properties of
language rather than the structure or contents of the stories
themselves. In this sense, social networks of characters and other
particular features of the narrative are not relevant for this
study. Instead, we are interested in the global organization of
words in the texts and in their networks. The structure of the
remainder of this chapter is as follows. In section~\ref{sec:2} we
start investigating the above-mentioned examples of the Ukrainian
language by analysing their frequency-rank dependencies. Verifying
that Zipf's law (\ref{eq1.2}) holds for the selected texts,
convinces us that the samples are large enough to deliver meaningful
statistical peculiarities. In section~\ref{sec:3} the details of how
to represent language by networks are described and the properties
of the corresponding language networks are investigated,  focusing
on the small-world and scale-free topologies. Conclusions are
outlined in section~\ref{sec:4}.

\section{Word appearance statistics}
\label{sec:2}

As mentioned above, the aim of our investigation is to perform a
quantitative analysis of the Ukrainian language, considering two
fables as its representatives. It is natural to start this analysis
by verifying the validity of Zipf's law (\ref{eq1.2}) for the texts
that represent each of the fables separately. In particular, this
will show whether the sizes of the selected texts are large enough
to demonstrate the expected statistical regularities.

The original part of our investigation uses electronic versions of
the two fables {\em Mykyta the Fox} and {\em Abu-Kasym's slippers}\footnote{The access
to the electronic versions of these texts was through
the most complete internet library of Ukrainian poetry,
\url{http://poetyka.uazone.net/}}, and is
based on our initial analysis of these texts (Holovatch and Palchykov 2007). All the
words were set to their canonical forms.
The lengths (total number of words) in these texts are $N=15426$ and $N= 8002$ for {\em Mykyta
the Fox} and {\em Abu-Kasym's slippers}, respectively.
The corresponding  vocabularies (the number of unique words) are $V=3563$ and $V= 2392$, respectively.
  The text comprising a combination of the two fables has $N=23428$ separate, and
$V=4823$ unique, words. Now, let us count the number of appearance
of each unique word in a text. We associate these numbers with the
frequencies of appearance $f$, even though they differ by the
normalization coefficient $N$. Ordering all the unique words by
decreasing frequencies of their appearance, one may assign the rank
variable $r=1,2,\ldots,V$ with each of them, such that the most
frequent word has a rank $r=1$, the rank of the second most frequent
word is $r=2$, etc. In cases where several unique words have exactly
the same frequencies of appearances, they are randomly assigned
sequential rank values without any preferences. The top ranked words
are the ones that appear the most frequently within the texts. For
the Ukrainian language these are function words that have little
semantic content of their own and chiefly indicate a grammatical
relationship. Words that are rather related to content of a
particular text tend to have lower frequencies. Two samples of such
ranking ordering are shown in Tab.~\ref{Table1}.
\begin{table}
\caption{Rank classification of words from Ivan Franko's {\em Mykyta
the Fox} (left part of the table) and {\em Abu-Kasym's slippers}
(right). The table shows some of the most frequently used words
(prepositions, pronouns, etc., as well as nouns)   in Ukrainian for
each of the two fables and their  English translations. Here $r$ is
the rank of the word and $f$ is the number of times it has appeared
in the text.} \label{Table1}
\begin{tabular}{p{0.7cm}p{1cm}p{2cm}p{1.5cm}|p{0.7cm}p{1cm}p{2cm}p{1.5cm}}
\hline\noalign{\smallskip}
$r$&    $f$&    word (in~Ukrainian)& English translation&   $r$&    $f$&    word (in~Ukrainian)&    English translation \\
\noalign{\smallskip}\svhline\noalign{\smallskip}
1&  439&    \foreignlanguage{ukrainian}{я}&     I&  1&  165&    \foreignlanguage{ukrainian}{він}&   he  \\
2&  323&    \foreignlanguage{ukrainian}{не}&    not&    2&  163&    \foreignlanguage{ukrainian}{в}&     in  \\
3&  312&    \foreignlanguage{ukrainian}{в}&     in& 3&  143&    \foreignlanguage{ukrainian}{не}&    not \\
4&  272&    \foreignlanguage{ukrainian}{і}&     and&    4&  140&    \foreignlanguage{ukrainian}{і}&     and \\
5&  233&    \foreignlanguage{ukrainian}{ти}&    you&    5&  128&    \foreignlanguage{ukrainian}{той}&   those   \\
6&  222&    \foreignlanguage{ukrainian}{що}&    that&   6&  125&    \foreignlanguage{ukrainian}{що}&    that    \\
7&  214&    \foreignlanguage{ukrainian}{на}&    on& 7&  125&    \foreignlanguage{ukrainian}{на}&    on  \\
$\vdots$&   $\vdots$&   $\vdots$&   $\vdots$&   $\vdots$&   $\vdots$&   $\vdots$&   $\vdots$    \\
16& 140&    \foreignlanguage{ukrainian}{лис}&   fox&    12& 87& \foreignlanguage{ukrainian}{капець}&    slipper \\
21& 109&    \foreignlanguage{ukrainian}{Микита}&    Mykyta& 18& 69& \foreignlanguage{ukrainian}{Абу-Касим}& Abu-Kasym\\
23& 98& \foreignlanguage{ukrainian}{вовк}&  wolf&   40& 28& \foreignlanguage{ukrainian}{пан}&   lord    \\
25& 88& \foreignlanguage{ukrainian}{цар}&   tsar&   41& 27& \foreignlanguage{ukrainian}{суддя}& judge   \\
\noalign{\smallskip}\hline\noalign{\smallskip}
\end{tabular}
\end{table}

\begin{figure}[b]
\begin{tabular}{cc}
\includegraphics[width=5.5cm]{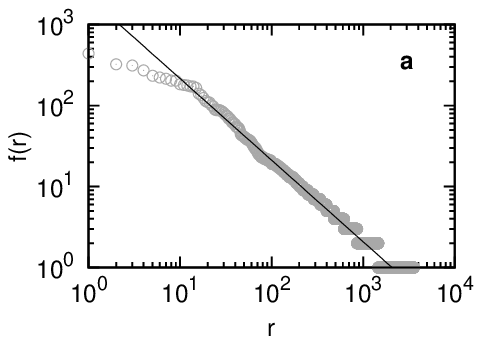} &
\includegraphics[width=5.5cm]{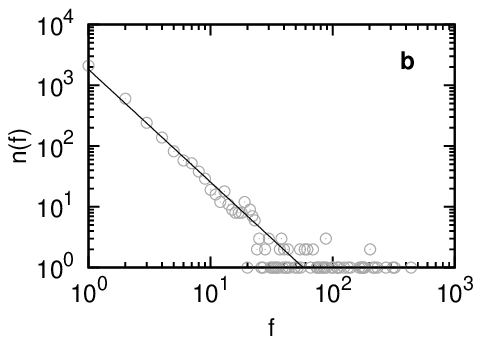} \\
\end{tabular}
\caption{Zipf laws for {\em Mykyta the Fox}.
Panel \textbf{a}: empirically observed frequency-rank dependency approximated by the power-law function (\ref{eq1.2}) with exponent $\alpha = 1.01$ for $r>20$.
Panel \textbf{b}: dependency of the number $n(f)$ of unique words  that appeared precisely $f$ times in
the text as a function of $f$, with the corresponding power-law
approximation from (\ref{eq2.3}) with $\beta = 1.85$.} \label{Figure1}
\end{figure}

The frequency-rank dependence for the words of {\em Mykyta the Fox} is shown in Fig.~\ref{Figure1}\textbf{a}.
This dependency may be accurately described by the power-law
function (\ref{eq1.2}) for  {roughly} $r>20$. The exponent of the
distribution has been estimated by using least-squares fitting with
a double logarithmic scale. (In cases where ranks have the same
frequencies $f$, a single $r$-value, namely its average on the
logarithmic scale, has been used for fitting purposes.) The
resulting exponents are close to the originally observed value;
$\alpha = 1.01\pm0.01$ for {\em Mykyta the Fox}, $\alpha =
0.99\pm0.02$ for {\em Abu-Kasym's slippers} and $\alpha =
1.00\pm0.01$ for the combined text of the two fables. Here the
accuracies of identification of exponent $\alpha$ are expressed by
asymptotic standard errors.

Let us stress one more specific feature of the dependency $f(r)$:
the number of unique words $n(f)$ that have the same frequency $f$
of appearance increases with $r$. The corresponding dependence for
{\em Mykyta the Fox} is shown in Fig.~\ref{Figure1}\textbf{b}. The
results of quantitative investigations (Zipf 1935, Zipf 1949) show
that this dependence also follows a power law decay:
\begin{equation}\label{eq2.3}
n(f) = B/f^{\beta},
\end{equation}
where $B$ is the proportionality coefficient. The reported values of
the exponent $\beta$ are close to $\beta=2$ (Zipf 1935), however,
some deviations from this value, caused by  specific lexicon
features, have been observed [see (Ferrer i Cancho 2005) and
references therein]. The values of the exponent $\beta$ for the
texts under our investigation were estimated to be
$\beta=1.85\pm0.04$ for {\em Mykyta the Fox}, $\beta=2.02\pm0.08$
for {\em Abu-Kasym's slippers} and $\beta=1.80\pm0.04$ for the
combination of the two texts. The fitting of $n(f)$ has been performed
in a similar manner to that of the $f(r)$ dependence: the most general
words with $r\leq20$ have been excluded and then for each value of
$n(f)$ a single value of $f$, its average in logarithmic scale, has
been assigned.

Eq. (\ref{eq2.3}) is sometimes referred to as the second Zipf
law [in which case Eq.(\ref{eq1.2}) is called the first].
However, the two  laws are not independent: equation (\ref{eq2.3})
  directly leads to (\ref{eq1.2}) asymptotically. To show this, let us
have a look at the rank of a word from a different point of view.
The rank $r$ of a word that has appeared exactly $f$ times may be {
{estimated by}} the number of unique words that have appeared  $f$
times or less:
\begin{equation}\label{eq2.3a}
r(f) = \sum_{f^{\prime}=f}^{f_{\rm max}} n(f^{\prime}),
\end{equation}
where $f_{\rm max}$ is the frequency of the most frequent word in
the text. Substituting (\ref{eq2.3}) into (\ref{eq2.3a}) and
assuming that $\beta>1$ one arrives at the inverse rank-frequency
dependency for infinitely large $f_{\rm max}$ as
\begin{equation}\label{eq2.3b}
r(f) = \frac{B}{\beta-1}f^{1-\beta}.
\end{equation}
Equation (\ref{eq2.3b}) matches equation (\ref{eq1.2}) provided that
\begin{equation}\label{eq2.3c}
\beta = 1+1/\alpha.
\end{equation}
Thus, $\alpha=1$ leads directly to $\beta=2$, as observed  empirically.

The power-law character of the frequency-rank distribution
(\ref{eq1.2}) may be explained by a number of theoretical
models.\footnote{The formation of a sentence may be considered
(Thurner et al. 2015) as an example of a history-dependent process
that becomes more constrained as it unfolds (Corominas-Murtra et al.
2015). Recently it has been demonstrated that stochastic processes
of this kind necessarily lead to Zipf's law too (Thurner et al.
2015, Corominas-Murtra et al. 2015).} One of the best known is a
generative Simon model (Simon 1955). This model belongs to a class
of models that are based on  so-called null hypotheses (Ferrer i
Cancho 2005). The null hypotheses ignore some fundamental aspects of
why and how the system units are used, but they often lead to
qualitatively correct descriptions of  system behaviour.
 The Simon model considers the process of text writing and uses two basic mechanisms to predict the $(n+1)^{\rm th}$ word
provided $n$ words have been already written and are known: i) the probability that the $(n+1)^{\rm th}$ word is one of the words that have appeared within the first $n$ words is proportional to its frequency
of appearance, and ii) there is a fixed probability $\delta$ that the $(n+1)^{\rm th}$ word will be a new one -- a word that has not
appeared within the first $n$ words.
Assuming that the text is generated accordingly to the Simon model, its frequency-rank dependence will asymptotically follow Zipf's law (\ref{eq1.2}) with  exponent (Simon 1955)
\begin{equation}
\alpha = 1-\delta.
\end{equation}
Even though the Simon model is not able to reproduce the subsequent
part of a text given its preceding part [due to its stochastic
origin and the existence of a number of words with the same
frequencies (\ref{eq2.3})], it was shown that its mechanism
plausibly describes  real writing processes (Holovatch and Palchykov
2007). However, to have a deeper understanding of the language
features, one has to go beyond Zipf's law that omits the
relationships between interacting words. Below we describe how the
structure of these relationships may be studied using the theory of
complex networks (Bornholdt and Schuster 2003).

\section{Language networks}
\label{sec:3}

The first step in applying complex network tools to investigate
quantitative properties of human language is to represent that
language as a network or a graph. Within such an interpretation  key
language units are considered as the nodes of the network and the
links reflect relations between them. Depending on the purpose of
investigation, different language units may be used as the network
nodes: phonemes, words, concepts or sentences being amongst the
possibilities. Then, focusing on one type of node, different
networks of language may be reconstructed for various
interpretations of the links that connect the nodes,  e.g. semantic
or syntactic relationships between words as described in the
Introduction.

\subsection{Network representations}
\label{subsec:3.1} 
\begin{figure}[b]
\centering
\includegraphics[width=11cm, natwidth=1454, natheight=784]{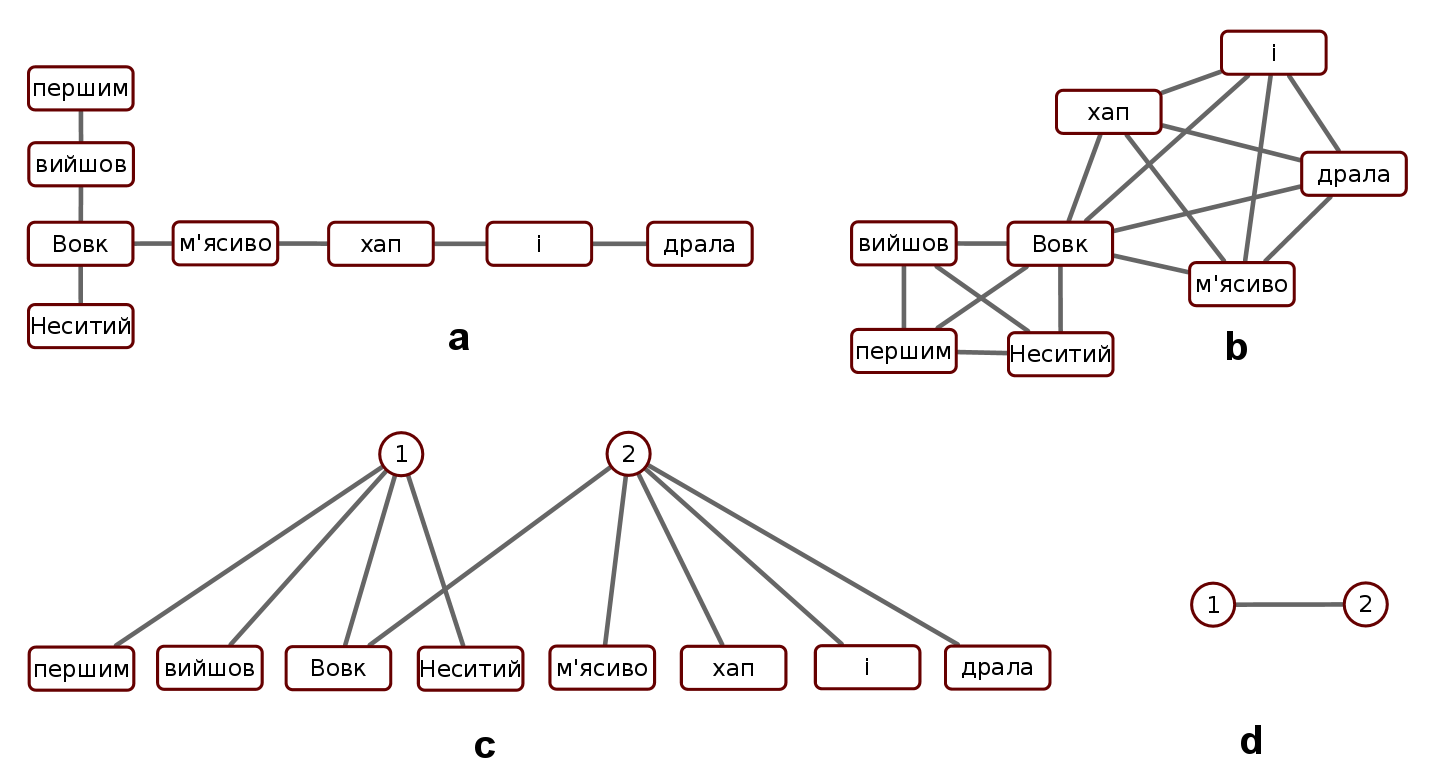}
\caption{Networked representation of the corpus of two sentences
that share a single common word
``\foreignlanguage{ukrainian}{вовк}'' (wolf). Panel \textbf{a}:
$L-space$ of human language. The co-occurring word-nodes are
connected by a link. Panel \textbf{b}: $P$-space. In this space the
links connect all word-nodes that belong to the same sentence.
Hence, each sentence appears in the network as a complete subgraph
(clique). Panel \textbf{c}: $B$-space. This space is represented by
a bipartite network that contains two types of nodes: word-nodes and
sentence-nodes. The links connect the nodes of different types only:
sentences and the words that have been found within. Panel
\textbf{d}: $C$-space. Here the nodes represent different sentences
and the two nodes are connected if they share at least a single
word.} \label{Figure3}
\end{figure}

In our investigation we analyse the properties of
language networks whose nodes correspond to unique words of a given
text. Let us connect a couple of nodes if the corresponding words
co-occur at least within a single sentence. We will refer to this
representation as the $L$-space of human language
(Fig.~\ref{Figure3}\textbf{a}). The links in $L$-space as well as in
the other spaces that will be described below\footnote{For different
network representations (different spaces) we use the nomenclature
originally introduced in the context of transportation networks
(Sienkiewicz and Ho\l yst 2005, von Ferber et al. 2007, von Ferber
et al. 2009).} are unweighted. Alternatively, the links may connect
not only the nearest neighbours, bur rather all words that are
located  a specific distance from each other. In order to take into
account this issue, we introduce a radius of interactions $R$: for
$R=1$ the links connect only co-occurring words (the nearest
neighbours), for $R=2$ the links connect the nearest and the next
nearest neighbours etc. Radius $R$ may be assigned any value in the
range $R\in[1, R_{\rm max}]$, where $R_{\rm max}+1$ is the size of
(the number of words within) the longest sentence. For the value
$R=1$ the resulting network reduces to the $L$-space of human
language. If $R=R_{\rm max}$ then the links connect all words that
belong to the same sentence and the corresponding network
representation will be referred to as the $P$-space of human
language (Fig.~\ref{Figure3}\textbf{b}).


An alternative way to represent a language network is to consider it as
a bipartite graph (Diestel 2005). We refer to this as $B$-space, see Fig.~\ref{Figure3}\textbf{c}. Such a representation
consists of two different types of nodes: the nodes of one type
represent sentences and the nodes of the other type represent words.
The link may connect the nodes of different types only and reflect the
appearance of given word within a specified sentence. Bipartite
networks provide a simple way to obtain two corresponding unipartite
networks by performing one-mode projections. One of the
resulting networks reproduces the $P$-space of human language. The
other one will be referred to as the language $C$-space
(Fig.~\ref{Figure3}\textbf{d}). In $C$-space the nodes represent
sentences and the link between two nodes  exist if the
corresponding sentences have at least one word in common.

The original part of our investigation will be focussed on the
networks of language whose nodes represent unique words. Varying the
values of $R$, will allow to perform a comparison
between the properties of different networks,
whose marginal realizations represent $L$- and $P$-spaces of
language.

\subsection{Basic network characteristics}
\label{subsec:3.2} To investigate the properties of the language networks
we will analyse a number of standard network characteristics:
the distribution $P(k)$ of node degrees, average node degree
$\langle{k}\rangle$, average clustering coefficient
$\langle{C}\rangle$ and the average value of the shortest path
length $\langle{l}\rangle$ between the nodes of the network that are
briefly described below.

The degree $k_i$ of a node $i$ is defined as the number of links
that connect that node with the other nodes of the network. The set
of all node degrees $\{k_i\}$ may be characterized by their
distribution $P(k)$: the probability that a randomly chosen node has
$k$ connections. Modifying the normalization condition one may
consider $P(k)$ as the number of nodes whose degree equals to $k$.
The distribution of node degrees contains all the necessary
information about fluctuations of the node degree around its average
value $\langle{k}\rangle$ that is defined as
\begin{equation}\label{eq3.20}
 \langle{k}\rangle = \frac{1}{V}\sum_{i=1}^V k_i.
\end{equation}
Here $i$ runs over all $V$ nodes of the network.
Both the average degree $\langle{k}\rangle$ and the shape of the degree distribution $P(k)$
are  key characteristics of the global network topology and its local fluctuations.

Networks with identical sequences of  degrees may vary significantly
due to the variations of the local connectivity patterns or
correlations. One kind of correlation, the local grouping of network
nodes, may be characterized by a clustering coefficient. The
clustering coefficient $C_i$ of  node $i$ is defined as the
probability that two randomly selected neighbours are connected with
each other:
\begin{equation}\label{eq3.21}
C_i = \frac{2m_i}{k_i(k_i-1)},
\end{equation}
where $m_i$ is the number of links that interconnect the
nearest neighbours of node $i$. The average value of clustering
coefficient
\begin{equation}\label{eq3.22}
\langle{C}\rangle = \frac{1}{V}\sum_{i=1}^V C_i
\end{equation}
characterizes the local grouping in the entire network and may be
used to compare the particular network with a random graph that
lacks  such correlations.

The shortest path length $l_{ij}$ between two nodes $i$ and $j$ is
defined as the minimal number of links that should be passed in
order to reach node $j$ starting at $i$. The average shortest path
length
\begin{equation}\label{eq3.23}
 \langle{l}\rangle = \frac{2}{V(V-1)}\sum_{i>j}l_{ij}
\end{equation}
is one of the characteristics of the network. Another
characteristics of the network is the maximal value of the shortest
path length $l_{\rm max} = {\rm max}(\{l_{ij}\})$.

\subsection{Ukrainian language networks}
\label{subsec:3.3}
Having introduced the main network
characteristics, let us investigate the properties of the language
networks of the selected fables. These characteristics for three
values of $R$: $R=1,2$ and $R_{\rm max}$ are summarized in
Tab.~\ref{tab:1}, and will be discussed in detail below.
\begin{table}
\caption{The basic quantitative characteristics of the investigated
language networks for several values of $R$. The upper part of the
table corresponds to the fable {\em Abu-Kasym's slippers}, the
middle part corresponds to {\em Mykyta the Fox} and the bottom part
represents the combined text of the two fables. The table contains
the number of nodes $V$; the number of links $M$; average  and
maximal node degrees $\langle{k}\rangle$ and $k_{\rm max}$
respectively; the exponents $\gamma$ and $\gamma_{\rm cum}$ of the
power-law fit to the degree distribution (\ref{eq1.1}) and
cumulative degree distribution (\ref{eq3.24}) respectively; the
average clustering coefficient $\langle{C}\rangle$ from
Eq.(\ref{eq3.22}) and  its counterpart $C_{\rm r}$ of the
corresponding random graph; and the average and maximal shortest
path lengths between nodes $\langle{l}\rangle$ and $l_{\rm max}$.}
\label{tab:1}
\begin{tabular}{p{1cm}p{1cm}p{1cm}p{1cm}p{1cm}p{1cm}p{1cm}p{1cm}p{1cm}p{1cm}p{0.5cm}}
\hline\noalign{\smallskip}
$R$&    $V$&    $M$& $\langle{k}\rangle$&   $k_{\rm max}$&  $\gamma$&   $\gamma_{\rm cum}$& $\langle{C}\rangle$&    $\langle{C}\rangle/C_{\rm r}$&  $\langle{l}\rangle$&    $l_{\rm max}$   \\
\noalign{\smallskip}\svhline\noalign{\smallskip}
1&          2392&   6273&   5.24&   228&    1.9&    1.2&   0.172&  78&  3.43&   11  \\
2&      2392&   11475&  9.59&   391&    2.0&    1.2&   0.567&  141& 2.90&   7   \\
$R_{\rm max}$&  2392&   48603&  40.64&  1134&   1.9&    1.4&   0.841&  50&  2.22&   4   \\
\noalign{\smallskip}\hline\noalign{\smallskip}
1&      3563&   11102&  6.23&   419&    1.9&    1.1&   0.214&  122& 3.30&   11  \\
2&      3563&   20063&  11.26&  665&    1.8&    1.2&   0.588&  186& 2.85&   7   \\
$R_{\rm max}$&  3563&   65997&  37.05&  1526&   1.9&    1.3&   0.822&  79&  2.27&   5   \\
\noalign{\smallskip}\hline\noalign{\smallskip}
1&      4823&   16580&  6.88&   537&    1.9&    1.1&   0.243&  170& 3.24&   11  \\
2&      4823&   29916&  12.41&  868&    1.8&    1.2&   0.585&  227& 2.83&   7   \\
$R_{\rm max}$&  4823&   107750& 44.68&  2185&   2.0&    1.3&   0.818&  88&  2.50&   5   \\
\noalign{\smallskip}\hline\noalign{\smallskip}
\end{tabular}
\end{table}

\begin{figure}[b]
\begin{tabular}{cc}
\includegraphics[width=5.5cm]{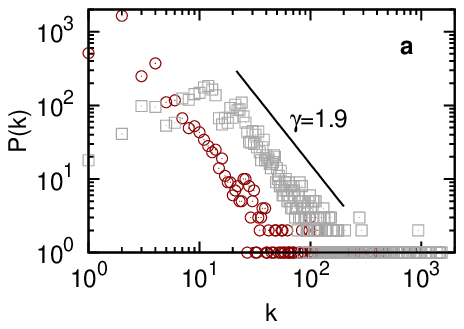} &
\includegraphics[width=5.5cm]{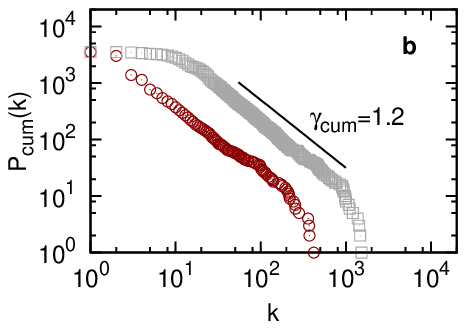} \\
\end{tabular}
\caption{{\em Mykyta the Fox}: node degree distributions (panel
\textbf{a}) and cumulative node degree distributions (panel
\textbf{b}) for $R=1$ (brown circles) and $R=R_{\rm max}$ (grey
squares). The solid lines are shown as the guides to the eye and
represent the power law decay functions (\ref{eq1.1}) in panel
\textbf{a} and (\ref{eq3.24}) in panel \textbf{b} with  exponents
$\gamma = 1.9$ and $\gamma_{\rm cum} = 1.2$, correspondingly.}
\label{Figure5}
\end{figure}

The distributions of the node degrees $P(k)$ for {\em Mykyta the
Fox} in $L$- and $P$-spaces, which give the number of nodes with
degree $k$, are shown in Fig.~\ref{Figure5}\textbf{a}.
Besides the peak around $k\sim10$, observed in $P$-space, the tails
of the distributions follow a straight line on a double logarithmic
scale, and the functional dependence of the node degree
distributions may therefore be described by the power law function.
Similar dependencies describe degree distributions of {\em
Abu-Kasym's slippers} and the combination of the two fables. The
exponent of the power law decay function (\ref{eq1.1}) fluctuates
around $\gamma=1.9$ to $2.0$, but the sizes of the investigated
networks do not allow one to make more precise estimations. In order
to justify the power law behaviour of the degree distribution, we
additionally consider the cumulative node degree distribution:
\begin{equation}\label{eq3.24}
P_{\rm cum}(k) = \sum_{k^\prime=k}^{k_{\rm max}} P(k^\prime).
\end{equation}
The corresponding dependencies for {\em Mykyta the Fox} are shown in
Fig.~\ref{Figure5}\textbf{b}. The function $P_{\rm cum}(k)$ is
smoother than $P(k)$ and allows one to make a direct conclusion on
the power-law behaviour of the degree distribution function.

A similar analysis for the English language has been performed for
the British National Corpus\footnote{The British National Corpus is
a collection of samples of written and spoken language from a wide
range of sources, designed to represent a wide cross-section of
British English from the late twentieth century,
\url{http://www.natcorp.ox.ac.uk/}.} with $V\sim10^7$. The analysis
of this corpus (Ferrer i Cancho and Sol\'e 2001b) demonstrates that
the corresponding language network is scale-free, with power-law
decay of the degree distribution { {$P(k)$}} characterized by two
distinct regimes with the exponent $\gamma=1.5$ for $k\leq2000$ and
$\gamma=2.7$ for $k\geq2000$. We cannot, of course, claim that the
values of the exponents which we have obtained for the two fables
persist for the entire corpus of the Ukrainian language. Nor can we
exclude the possibility of a crossover in the entire corpus of the
type  observed in (Ferrer i Cancho and Sol\'e 2001b). However, the
results of this pilot study demonstrate that the network of the
Ukrainian language used, at least for the two fables analysed,  is
characterized by a scale-free topology.

Besides exhibiting  scale-free topologies, many real networks   tend
to be small worlds (Albert and Barab\'asi 2002, Watts 1999). A
network is considered to be a small world if its average shortest
path length $\langle{l}\rangle$ increases with the number of nodes
$V$ slower that any power-law function (Dorogovtsev and Mendes
2003). Note for comparison that a regular $d$-dimensional lattice
has $\langle{l}\rangle\sim V^{1/d}$. Small worlds are extremely
compact; an arbitrary pair of nodes is separated just by a few
links. The notion is known in sociology (where it originated), where
it has been shown that two randomly chosen members of society are
separated by an average of six intermediate relationships (Milgram
1967). Tab.~\ref{Table1} shows that for $R=1$ the maximal shortest
path length for all three networks  is $l_{\rm max}=11$ and the
average shortest path length is $\langle{l}\rangle\sim3$. The
average $\langle{l}\rangle$ and the maximal $l_{\rm max}$ decreases
with $R$, reaching about 4 or 5 for $l_{\rm max}$ and just above 2
for $\langle{l}\rangle$ in $P$-space. Such behaviour is quite
natural, since the number of links may only increase (or remain
unchanged) with increasing $R$ without affecting the number of nodes
$V$. This extreme compactness suggests that the Ukrainian language
networks used in the fables are characterized by the small-world
effect, even though a strict conclusion would require a study of
size dependent evolution. For comparison, the average shortest path
length $\langle{l}\rangle$ for the above-mentioned
 English resource is  $\langle{l}\rangle=2.63$ (Ferrer i Cancho and Sol\'e 2001b).

Connected triangles of nodes are typical signs for the presence of
correlations in networks. Defined in Eq.(\ref{eq3.22}), the average
clustering coefficient $\langle{C}\rangle$ is expected to
characterize this type of correlation. The clustering coefficient
for a complete graph is $\langle{C}\rangle=1$ and it is
$\langle{C}\rangle=0$ for a tree-like network. To characterize the
level of these local correlations, the clustering coefficient of a
network is usually compared to the one of a random graph with the
same number of nodes $V$ and links $M$, for which
\begin{equation}\label{eq3.25}
C_r = \frac{2M}{V^2}.
\end{equation}
Table~\ref{tab:1} gives  average values of the clustering coefficients
$\langle{C}\rangle$ and their ratios to the those for the corresponding random graphs
$C_r$. Since the observed clustering coefficients $\langle{C}\rangle$
 vastly exceed their random counterparts $C_r$, the
networks under consideration are well correlated structures. As
expected, these correlations become stronger with the radius of
interaction $R$, see  {Tab.~\ref{tab:1}}.

\begin{figure}[b]
\begin{tabular}{cc}
\includegraphics[width=5.5cm]{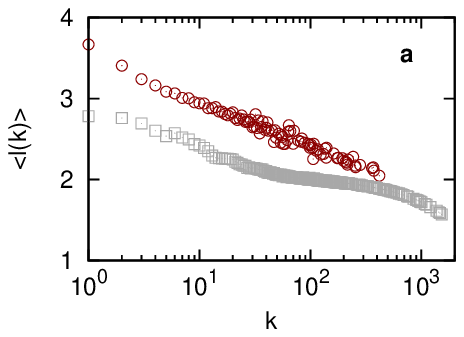} &
\includegraphics[width=5.5cm]{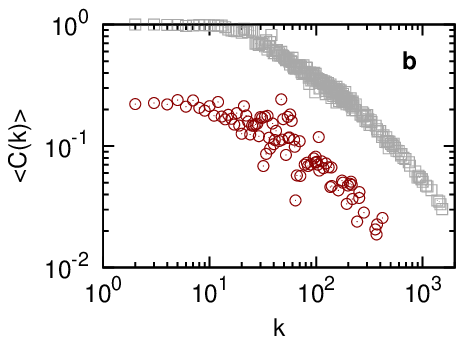} \\
\end{tabular}
\caption{{\em Mykyta the Fox}: average shortest path length
$\langle{l(k)}\rangle$ that separates a node of degree $k$ from the
other nodes of the network as a function of $k$ (panel \textbf{a}),
and the dependence of the average clustering coefficient
$\langle{C(k)}\rangle$ of a node with $k$ neighbours on $k$. The
$L$- ($R=1$) and $P$-spaces ($R=R_{\rm max}$) are represented by
dark circles and grey squares, respectively. } \label{Figure7}
\end{figure}

Finally, let us investigate the influence of the node degree $k$ on
the distance that separates the node from the rest of the network
and on the connectivity patterns of its neighbourhood.
The first of these (distance) is quantified by the average shortest path length
$\langle{l(k)}\rangle$ from a node of degree $k$ to an arbitrary chosen reachable
network node.
The connectivity patterns of the neighboring nodes are
described by the average clustering coefficient $\langle{C(k)}\rangle$ of
the nodes of degree $k$.
Fig.~\ref{Figure7}\textbf{a} shows that the average distance
$\langle{l(k)}\rangle$ from a node of degree $k$ to a randomly
selected other node of the network monotonically decreases with $k$.
This reflects the observation that the higher the degree of the
node, the less distance separates the node from the rest of the
network.
This short-distance scenario for the hub nodes is complemented by
the second scenario: the more neighbours the node has, the less
connected these neighbours are amongst each other, see
Fig.~\ref{Figure7}\textbf{b}. A high level of clustering coefficient
is observed only for the nodes with small degree, and it decreases
rapidly for  higher degree nodes.

It is interesting to compare the results of our analysis with
studies of collections of texts written in Portuguese and English,
whose sizes ranged between 169 and 276425 words (Caldeira et al.
2006). The representation of texts used in (Caldeira et al. 2006)
corresponds to $P$-space in our classification. However, unlike our
investigation, a network of concepts has also been considered in
(Caldeira et al. 2006) and could form the basis for a future study
of fables. The quantitative characteristics of the corresponding
network are $\gamma = 1.6\pm0.2$, $\langle{l}\rangle=2.0\pm0.1$,
$l_{\rm max} = 4\pm1$, $\langle{C}\rangle=0.83\pm0.03$ and are in a
good agreement with our results ($\gamma = 2.0$,
$\langle{l}\rangle=2.25$, $l_{\rm max} = 5$,
$\langle{C}\rangle=0.818$ for the combination of two fables). Such
agreement is interesting, not only because the investigated networks
correspond to different languages (English, Portuguese in Caldeira
et al. (2006) and Ukrainian in our case), but mainly because it
shows that the restriction to the specific subset of words
(words-concepts) does not cause significant changes in the
above-considered  features of a language network.

\section{Conclusions}
\label{sec:4} In this article, a quantitative analysis of the word
distribution in two fables written in Ukrainian by Ivan Franko ({\em
Mykyta the Fox} and {\em Abu-Kasym's slippers}) has been performed.
Our investigation consists of two distinct parts: analysis of the
frequency-rank dependence (Zipf's law) and a deeper analysis of the
structure of language using the tools developed within complex
network science. A main purpose of analyzing the frequency-rank
dependencies is to verify whether the texts under investigation are
large enough to exhibit the expected statistical features. Having
confirmed the validity of Zipf's law for the rank values in the
range $r=20-3000$, we have justified that these texts may be used to
investigate deeper features of the Ukrainian language used in those
fables.

The results of our investigation confirm that the network of the
Ukrainian language used is a highly correlated, small world that is
characterized by a scale-free topology. This is quite expected,
since similar results have been formerly obtained for various other
languages (Ferrer i Cancho and Sol\'e 2001b, Caldeira et al. 2006,
Zhou et al. 2008). The small-world effect highlights an extremely
high level of compactness of such networks: despite the large size
of the vocabulary used, a pair of randomly selected words is
separated on average by only three steps. The high values for
clustering coefficients observed in our analysis reflect the high
level of correlations in the network structures. The  empirical
results obtained here may be used for a theoretical description of
the evolution of language, which may be based on  evolutionary game
theory (Nowak and Krakauer 1999).

Some attempts have been made to explain  features of language
networks using the preferential attachment scenario (Albert et al.
1999); one attempts to consider such networks as the results of a
growth process, wherein  new words that join the network tend to be
connected to  hubs with higher probability than to  low degree nodes
(Ferrer i Cancho and Sol\'e 2001b, Caldeira et al. 2006, Dorogovtsev
and Mendes 2001). It is worth noting that, following such an
approach, the emergence of syntax is a consequence of the evolution
of language (Ferrer i Cancho et al. 2005, Sol\'e 2005).

Our  analysis of language networks in different spaces shows that
basic features, such as small-world effect or scale-free topology,
are space independent and highlight the properties of the language
itself rather than of a particular representation. This is strongly
confirmed by our comparison of distinct networks in $P$-space, which
shows that the restriction towards some categories of words does not
cause significant changes in the features of these networks.

In this chapter we report on an attempt to analyze some properties
of the Ukrainian language based on two narratives. On the one hand,
indeed one has to take a larger database to study the entire
language. On the other hand, as we show in section \ref{sec:2}, the
universal properties of word distribution already hold for our
database. It gives hope that some universalities of language can be
checked using this database too. Moreover, in similar studies texts
of similar word numbers have been used, see e.g. Caldeira et al.
(2006). An interesting question for further analysis might be to
search the non-universal, specific properties of the language used
in a given narrative or rather in a given group of narratives of a
similar type. Could one identify (or even categorise) genres through
such properties? A question of interest for the main readership (and
authors) of this book probably would be whether quantitative
features of the language of mythological narratives make them
different from other texts? We think that analysis performed from
such perspectives may lead to better understandings of what is still
hidden in old narratives. In other words, this is a first step of a
new programme and there is much more to do.

\begin{acknowledgement}
It is our pleasure to thank the Editors of this book Ralph Kenna,
M\'air\'in Mac Carron, and P\'adraig Mac Carron for their invitation
to contribute and for their help and discussions during preparation
of the manuscript. Yu.H. acknowledges useful discussions with Bernat
Corominas-Murtra. This work was supported in part by the 7th FP,
IRSES projects No.~295302 Statistical Physics in Diverse
Realizations (SPIDER), No.~612707 Dynamics of and in Complex Systems
(DIONICOS), by the COST Action TD1210 Analyzing the dynamics of
information and knowledge landscapes (KNOWSCAPE) and by SNSF project
No.~147609 Crowdsourced conceptualization of complex scientific
knowledge and discovery of discoveries.
\end{acknowledgement}

\vspace{1cm}


\end{document}